%% file: supix1_mainTex.tex
\documentclass[a4paper,11pt]{article}
\usepackage{lineno}% \linenumbers to turn on
\usepackage{amsmath}% after {lineno}
\usepackage{etoolbox} %% <- for \cspreto, \csappto
\newcommand*\linenomathpatch[1]{%
  \cspreto{#1}{\linenomath}%
  \cspreto{#1*}{\linenomath}%
  \csappto{end#1}{\endlinenomath}%
  \csappto{end#1*}{\endlinenomath}%
}
\linenomathpatch{equation}

\usepackage{jinstpub} % for details on the use of the package, please see the JINST-author-manual
\hypersetup{breaklinks}
\graphicspath{{./}{figure/}} %wm% searching path

\usepackage{caption}
\usepackage{subfigure}
\usepackage{multirow}

\usepackage{siunitx}
\newcommand*{\refcite}[1]{ref.~\cite{#1}}
\newcommand*{\tab}[1]{Table~\ref{tab:#1}}
\newcommand*{\fig}[1]{Figure~\ref{fig:#1}}
\newcommand*{\sect}[1]{Section~\ref{sect:#1}}
\newcommand*{\eq}[1]{eq.~(\ref{eq:#1})}
\newcommand*{\X}[3][]{\ensuremath{{}_{#1}^{#3}\text{{#2}}}\xspace}% \X[Z]{name}{A} element
\newcommand*{\mean}[1]{\left\langle#1\right\rangle}
\newcommand*{\supixi}{\textsc{Supix}-1\xspace}

\title{\boldmath Characterization of a CMOS pixel sensor for charged
  particle tracking }

\author[a]{L. Li,}
\author[a, 1]{L. Zhang, \note{Corresponding author.}}
\author[a]{J. N. Dong,}
\author[b]{J. Liu,}
\author[a, 1]{and M. Wang}

\affiliation[a]{ Institute of Frontier and Interdisciplinary Science
  and Key Laboratory of Particle Physics and Particle Irradiation of
  Ministry of Education, Shandong University, Qingdao 266237, China}

\affiliation[b]{Department of Physics, University of Liverpool,
  Liverpool, L693BX, United Kingdom}

\emailAdd{mwang@sdu.edu.cn}
\emailAdd{zhang.l@sdu.edu.cn}

\abstract{
  A prototype of the CMOS pixel sensor named \supixi has been
  fabricated and tested in order to investigate the feasibility of a
  pixelated tracker for a proposed Higgs factory, namely, the Circular
  Electron-Positron Collider (CEPC).
  The sensor, taped out with a 180\nm CMOS Image Sensor (CIS)
  process, consists of nine different pixel arrays varying in pixel
  pitches, diode sizes and geometries in order to study the particle
  detection performance of enlarged pixels.
  The test was carried out with a \X{Fe}{55} radioactive source. Two
  soft X-ray peaks observed were used to calibrate the charge to
  voltage factor of the sensor. The pixel-wise equivalent noise
  charge, charge collection efficiency and signal-to-noise ratio were
  evaluated.
  A reconstruction method for clustering pixels of a signal has been
  developed and the cluster-wise performance was studied as well.
  The test results show that pixels with the area as large as of
  \SI{21x84}{\um} have satisfactory noise level and charge collection
  performance, meeting general requirements for a pixel sensor.
  This contribution demonstrates that the CMOS pixel sensor with
  enlarged pitches, using the CIS technology, can be used in tracking
  for upcoming collider detectors akin to the CEPC.
}


\arxivnumber{2111.00290} % only if you have one

\begin{document}
\maketitle
\flushbottom

\section{Introduction}
\label{sect:a}

%% 研究背景、研究目的及技术要求

The discovery of the Higgs boson~\cite{higgsatlas, higgscms}
opens a new window for studying the fundamental physics principles of
the Standard Model (SM) and exploring beyond the SM.
The Circular Electron-Positron Collider (CEPC)~\cite{cepcacc} is a
proposed Higgs factory to conduct precision measurements of the Higgs
boson.
In the conceptual design of the CEPC detector~\cite{cepcphys}, the
tracking system involves a silicon tracker, with silicon microstrip
sensors as the basic technology. However, a fully pixelated silicon
tracker with large pixels is also under consideration and the CMOS
Pixel Sensor (CPS) has gained particular interest due to high
granularity and low material budget.
%
% physics requirements
The CEPC high precision tracking requires a general condition of the
single point resolution, $\sigma_\text{sp} < 7\um$, on the transverse
plane with regard to the magnetic field. In the longitudinal
direction, however, the resolution is undefined, but a loose
requirement at least a few times larger than $7\um$ is
anticipated. Without considering the effect of charge sharing, the
nominal pixel size will be about \SI{20x80}{\um} or larger.
With the standard CMOS Image Sensor (CIS) process, the CPS pixels
realized so far used in high-energy physics experiments are relatively
small.  For instance, a CPS based vertex detector for the STAR
experiment at the Relativistic Heavy Ion Collider (RHIC), Brookhaven
National Laboratory, employs pixels of
\SI{20.7x20.7}{\um}~\cite{starPXL}.  ALPIDE, another CPS chip which is
going to be used for the major upgrade of the inner tracking system
for the ALICE experiment at Large Hadron Collider (LHC), CERN, is
developed in \SI{27x29}{\um} pixels~\cite{alpide}.

\begin{figure}[htb!]
  \centering
  \subfigure[]{
    \includegraphics[width=0.99\textwidth,origin=c]{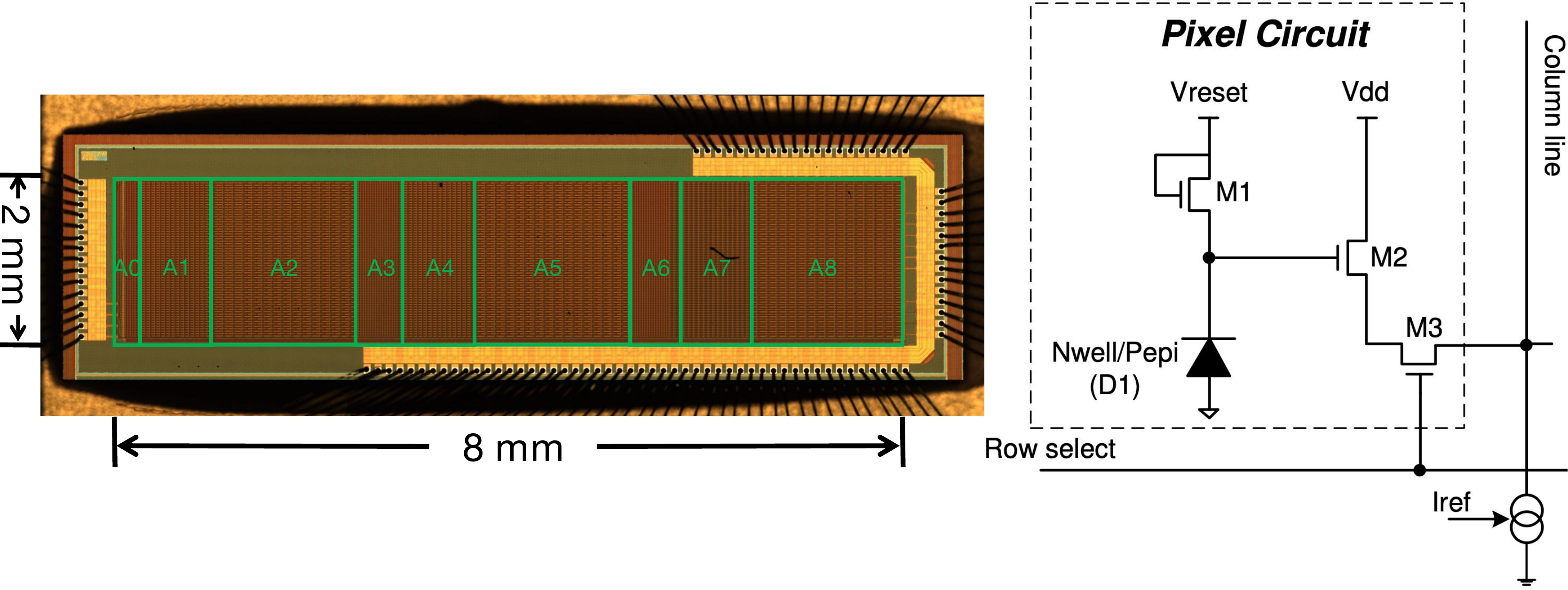}
    \label{fig:chip:a}
  }
  
  \subfigure[]{
    \includegraphics[width=0.49\textwidth]{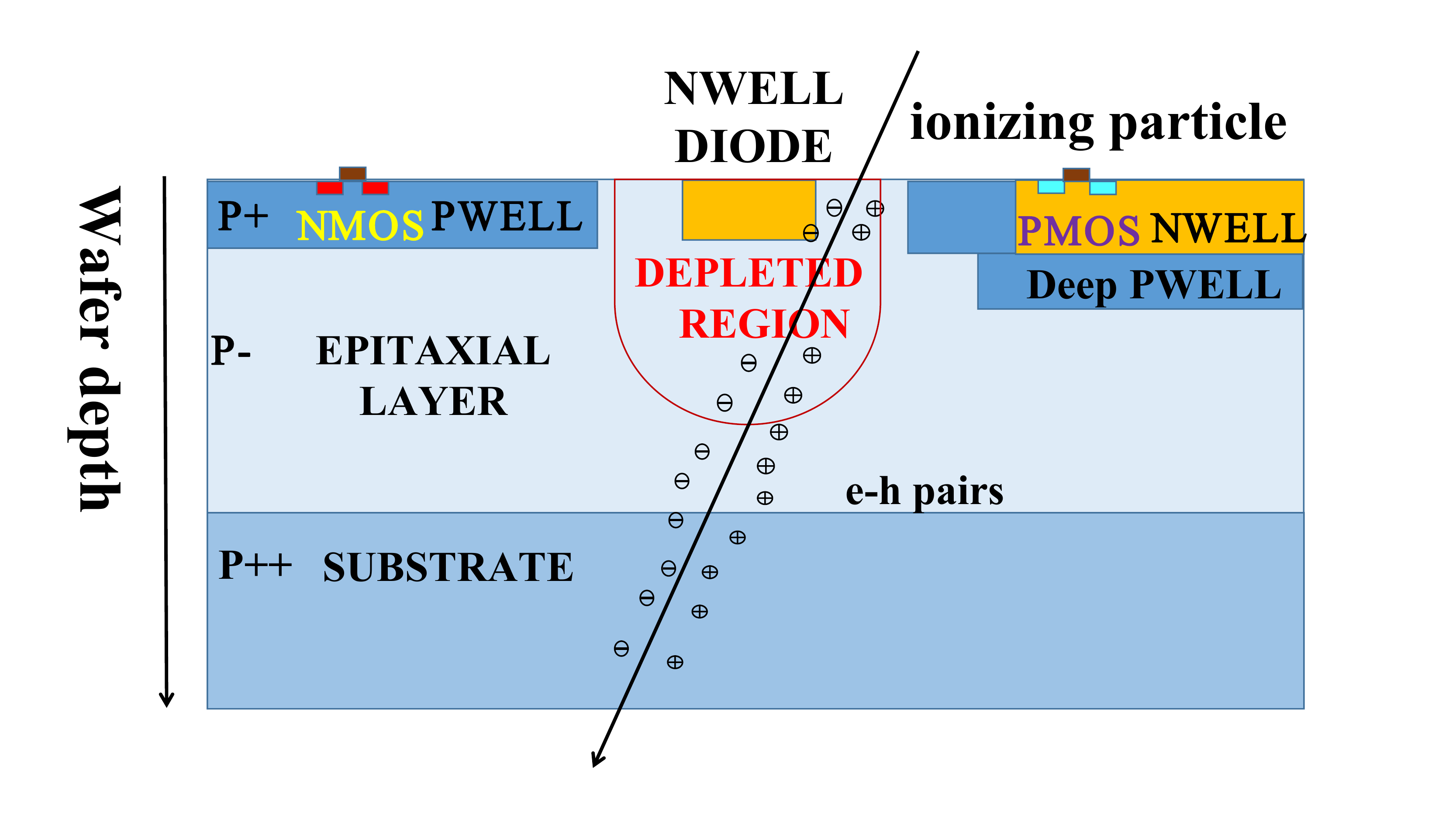}
    \label{fig:chip:b}
  }%
  \subfigure[]{
    \includegraphics[width=0.49\textwidth]{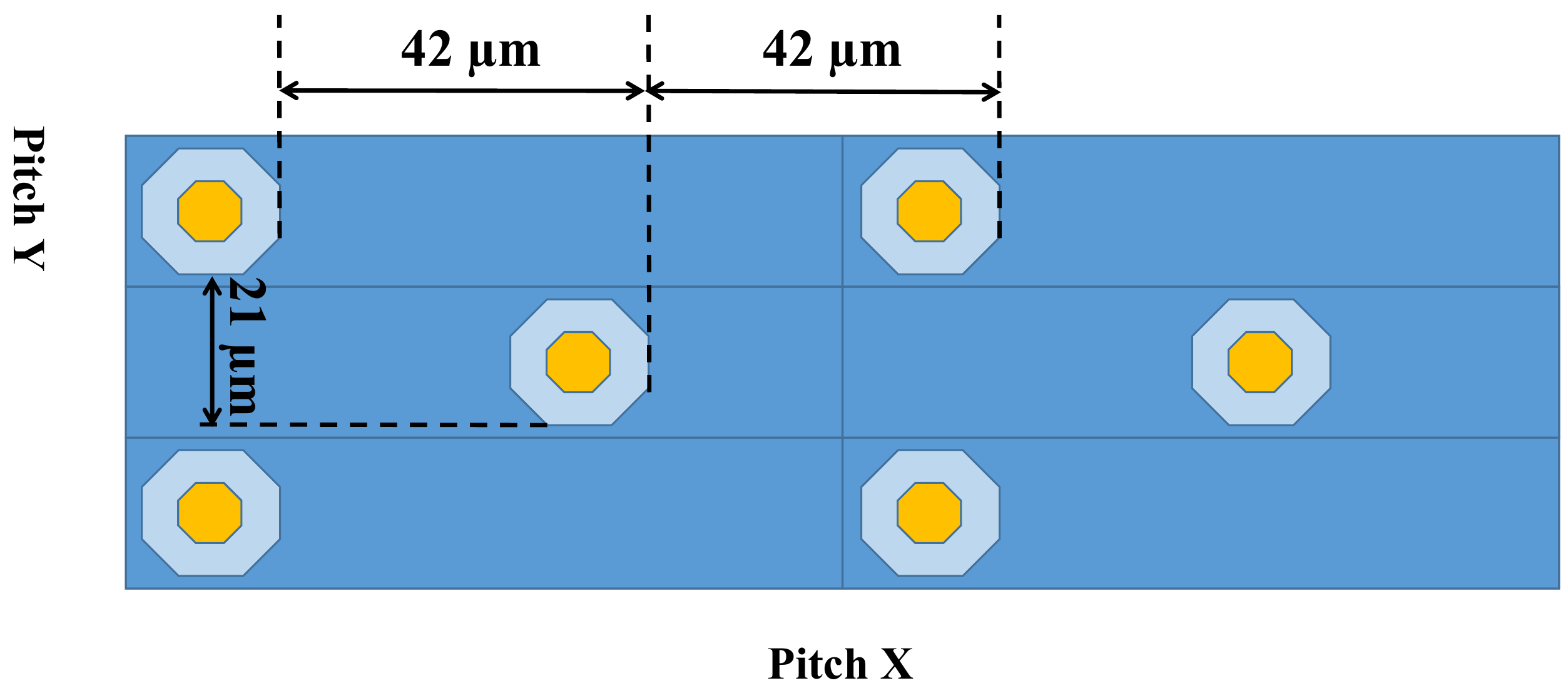}
    \label{fig:chip:c}
  }
  
  \caption{Design of the \supixi sensor: (a) a photograph and the
    schematic of in-pixel circuit, (b) illustration of a CMOS pixel
    cross section and (c) the staggered arrangement of diodes in
    sectors with the largest x-pitch.
  }
  \label{fig:chip}
\end{figure}

In order to investigate the feasibility of relatively large pixels
based on the standard CMOS imaging process, a prototype chip, named
\supixi (Shandong University PIXel), has been fabricated and tested.
The chip is designed with three different pixel sizes, up to
\SI{21x84}{\um}, and variant diode geometries. The main purpose of the
design is to explore the interplay of the pixel pitch and the diode
geometry on the charge collection efficiency.
Details of the chip design can be found in \refcite{supix1tcad}, while
some characteristics are briefed here.
As shown in \fig{chip:a}, the chip contains nine sectors (only sectors
A0, A2, A5, A7 and A8 were accessible during the test) and has an
overall area of \SI{2x8}{mm}.  Each sector contains 64 rows and 16
columns (labelled from C0 to C15) of pixels. The sensitive areas and the
pixel pitches of relevant sectors are listed in \tab{result}.
The schematic of in-pixel circuit is also shown in \fig{chip:a},
employing a 3T architecture. The collected charge in a pixel is
converted into the signal voltage through an N-well/P-epitaxial layer
diode~(D1), as illustrated in \fig{chip:b}.
Combinations of various diode geometries and pixel pitches are
implemented in different sectors, as listed in \tab{result}. The  diode
surface is defined as the area of the N-well and the diode footprint is
the total area formed by the surrounding P-well.
A staggered pixel design is implemented in sectors with the largest x-pitch
(\SI{84}{\um}), as shown in \fig{chip:c}, to enhance charge collection.
The analog signal of each sector is read out in a rolling shutter mode.
The readout time for a frame is \SI{32}{\us} with a basic clock of
\SI{2}{MHz}.

In this article, we present the test of \supixi with a radioactive
source of \X{Fe}{55}.
The procedure for characterizing the sensor performance, taking the
sector of A0 as an example, is described.
% The performance of pixels with different structures is discussed.
The manuscript is organized as follows:
a brief introduction of the test system is given in \sect{system},
steps of the test are elaborated in \sect{test}, test results are
discussed in \sect{results},
and we conclude in \sect{conclusion}.

\section{The test system for \supixi}
\label{sect:system}

\begin{figure}[htb!] 
  \centering
  \subfigure[]{
    \includegraphics[height=0.18\textheight]{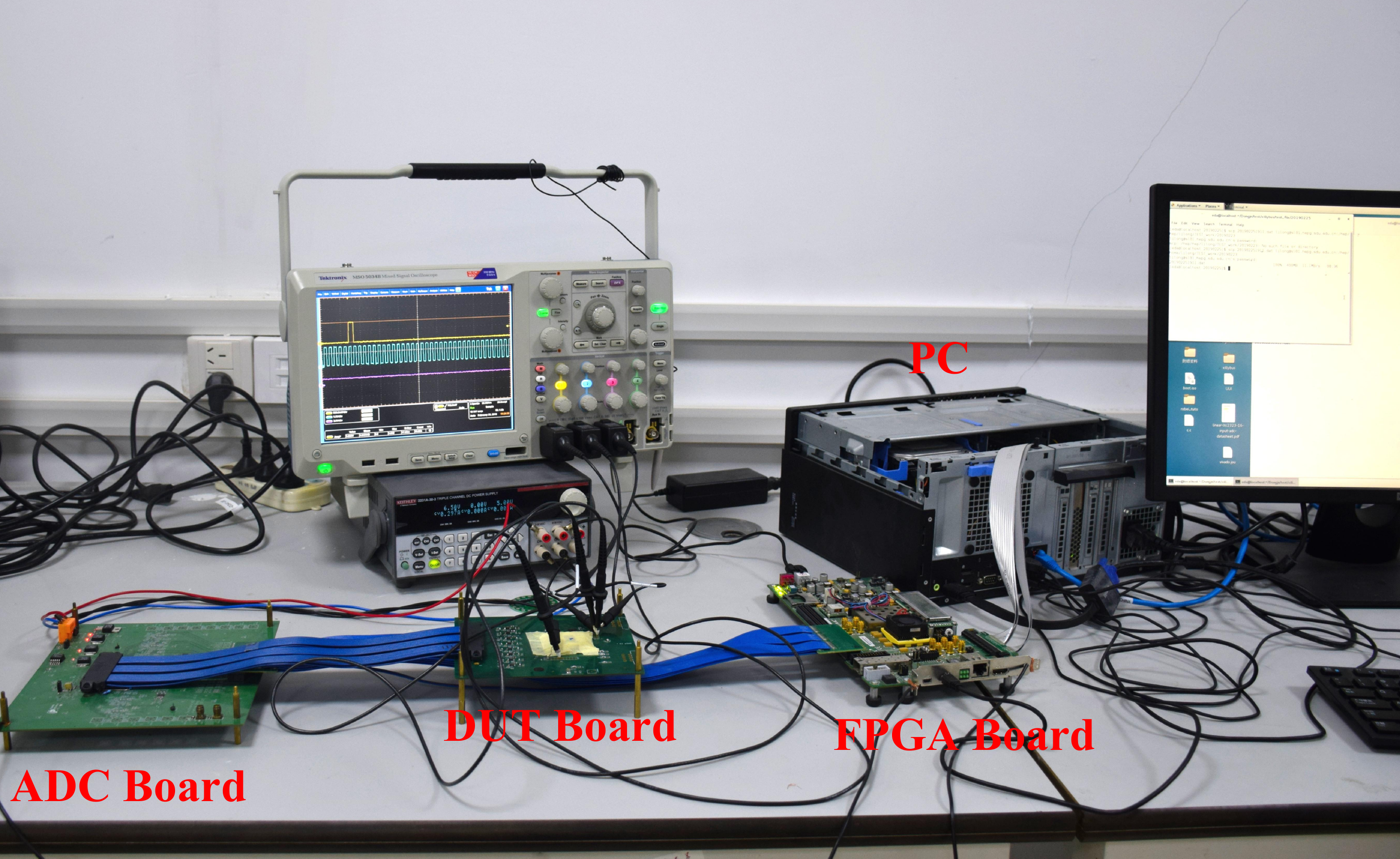}
    \label{fig:edaq:a}
  }~%
  \subfigure[]{
    \includegraphics[height=0.18\textheight]{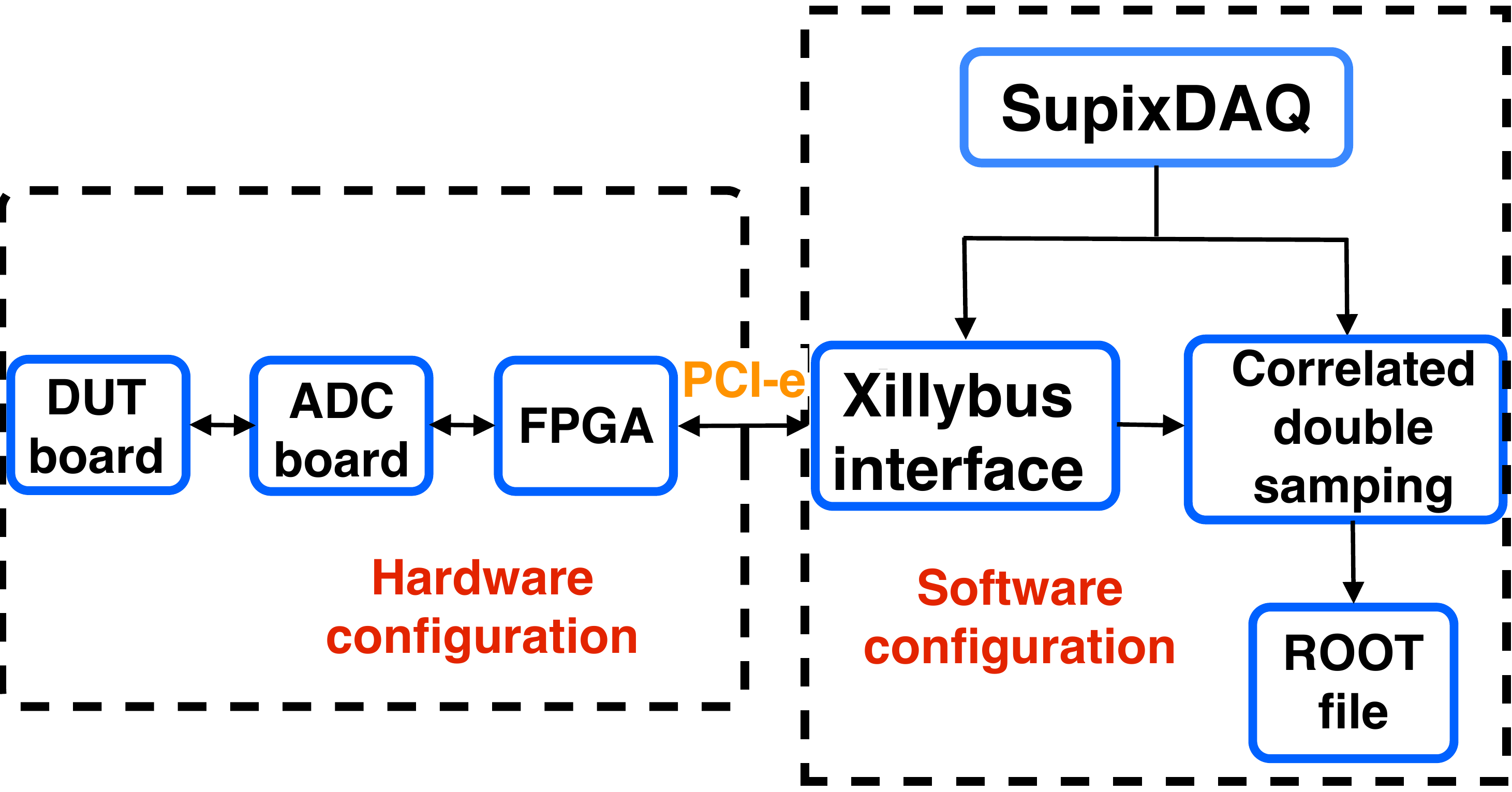}
    % "\includegraphics" from the "graphicx" permits to crop (trim+clip)
    % and rotate (angle) and image (and much more)
    \label{fig:edaq:b}
  }

  \caption{(a) Test system for the \supixi sensor and (b) working flow
    of the system.}
  \label{fig:edaq}
\end{figure}

As shown in \fig{edaq}, the test system for \supixi consists of
readout electronics and a data acquisition system (SupixDAQ).
The readout electronics involves a device under test (DUT) board, an
analog-to-digital conversion (ADC) board, a field programmable gate
array (FPGA) board as well as a personal computer (PC).
\begin{description}
\item [The DUT board] connects to the \supixi chip by wire bonding and
  drives analog signal out of the chip.
\item [The ADC board] digitizes the analog signal with 16-bit ADCs
  (LTC2323-16) and transmits digitized data to the FPGA board through
  a FPGA mezzanine card (FMC).
\item [The FPGA board] configures the ADC board and the DUT board and
  transmits the digitized data to the PC for data taking through a
  PCI-express interface. The FPGA is Xilinx Kintex-7 FPGA KC-705.
\item [The PC] manages data acquisition and records the data.
\end{description}

\begin{figure}[htb!] 
  \centering
  \subfigure[]{
    % \centering % \begin{center}/\end{center} takes some additional vertical space
    \includegraphics[width=0.49\textwidth,origin=c]{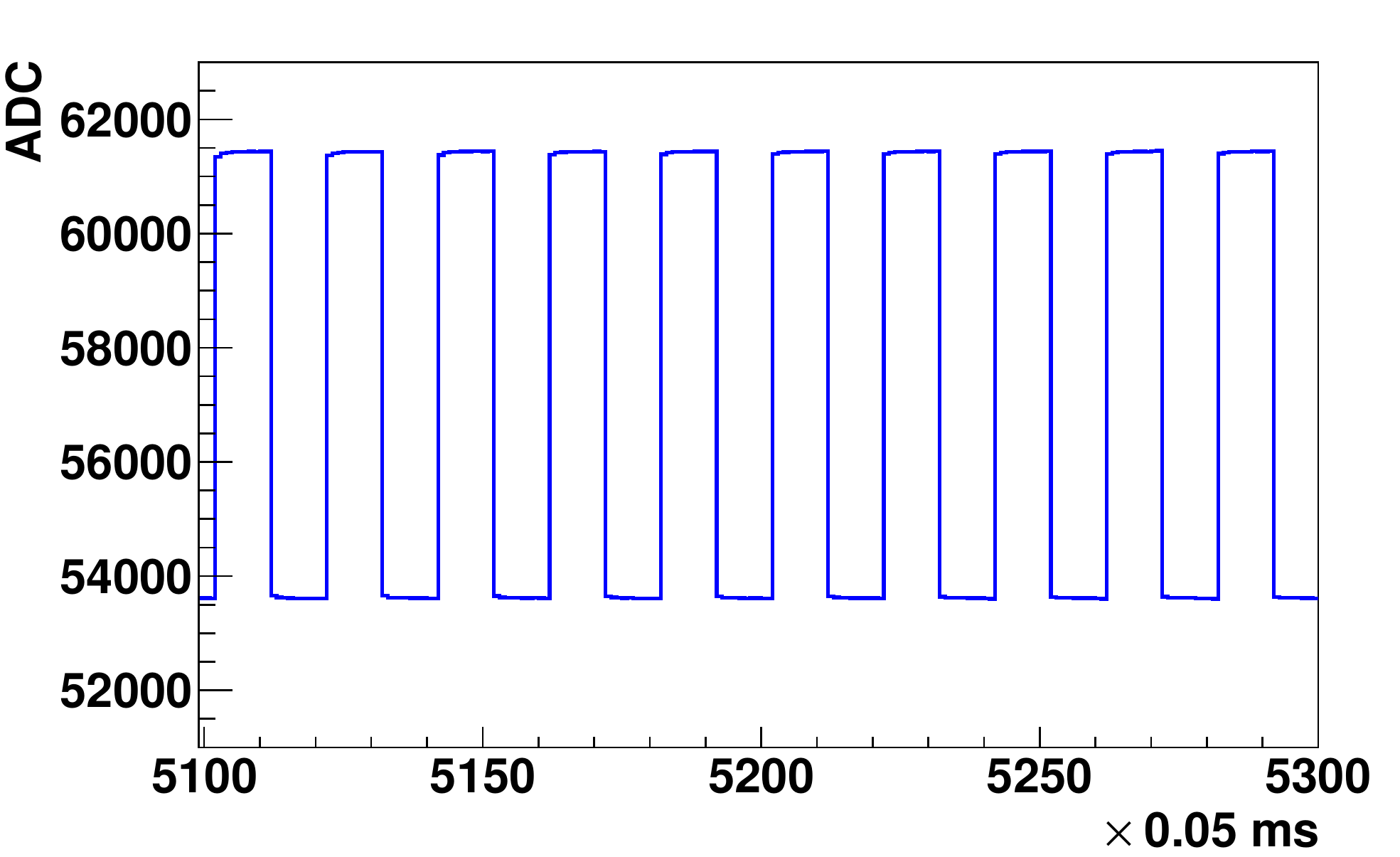}
    \label{fig:adc:a}
  }~%
  \subfigure[]{
    \includegraphics[width=.49\textwidth,origin=c]{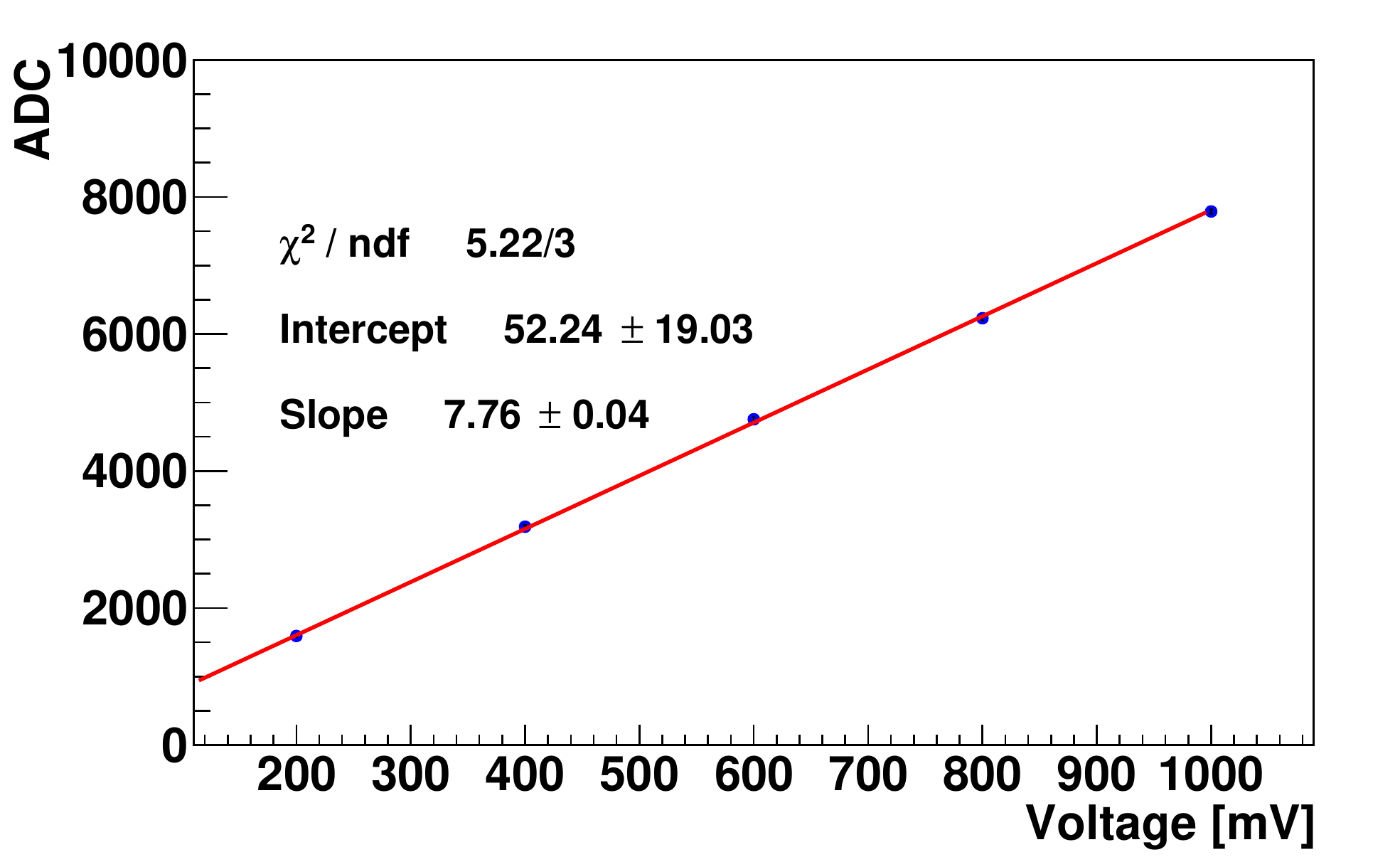}
    % "\includegraphics" from the "graphicx" permits to crop (trim+clip)
    % and rotate (angle) and image (and much more)
    \label{fig:adc:b}
  }

  \caption{Calibration for the readout electronics: (a) the ADC output 
    of an input square wave with \SI{1}{V} amplitude and 1\kHz
    frequency and (b) the ADC output as a function of the input voltage
    together with a linear function fit.}
  \label{fig:adc}
\end{figure}

The readout electronics was calibrated with the input voltage from a
signal generator (Tektronix AFG3252C). As shown in \fig{adc}, the
linearity is good in the range of the input voltage.
The electronics calibration constant is defined as
\begin{equation}
  \label{eq:electronics}
  A \equiv \frac{\text{ADC}}{\text{Voltage}},
\end{equation}
where ADC represents the ADC count. % (same in the following).
The calibration resulted in $A = \SI{7.76+-0.04}{ADC/mV}$.

%% CDS

\begin{figure}[htb!] 
  \centering
  \subfigure[]{
   % \begin{center}/\end{center} takes some additional vertical space
    \includegraphics[width=.32\textwidth,origin=c]{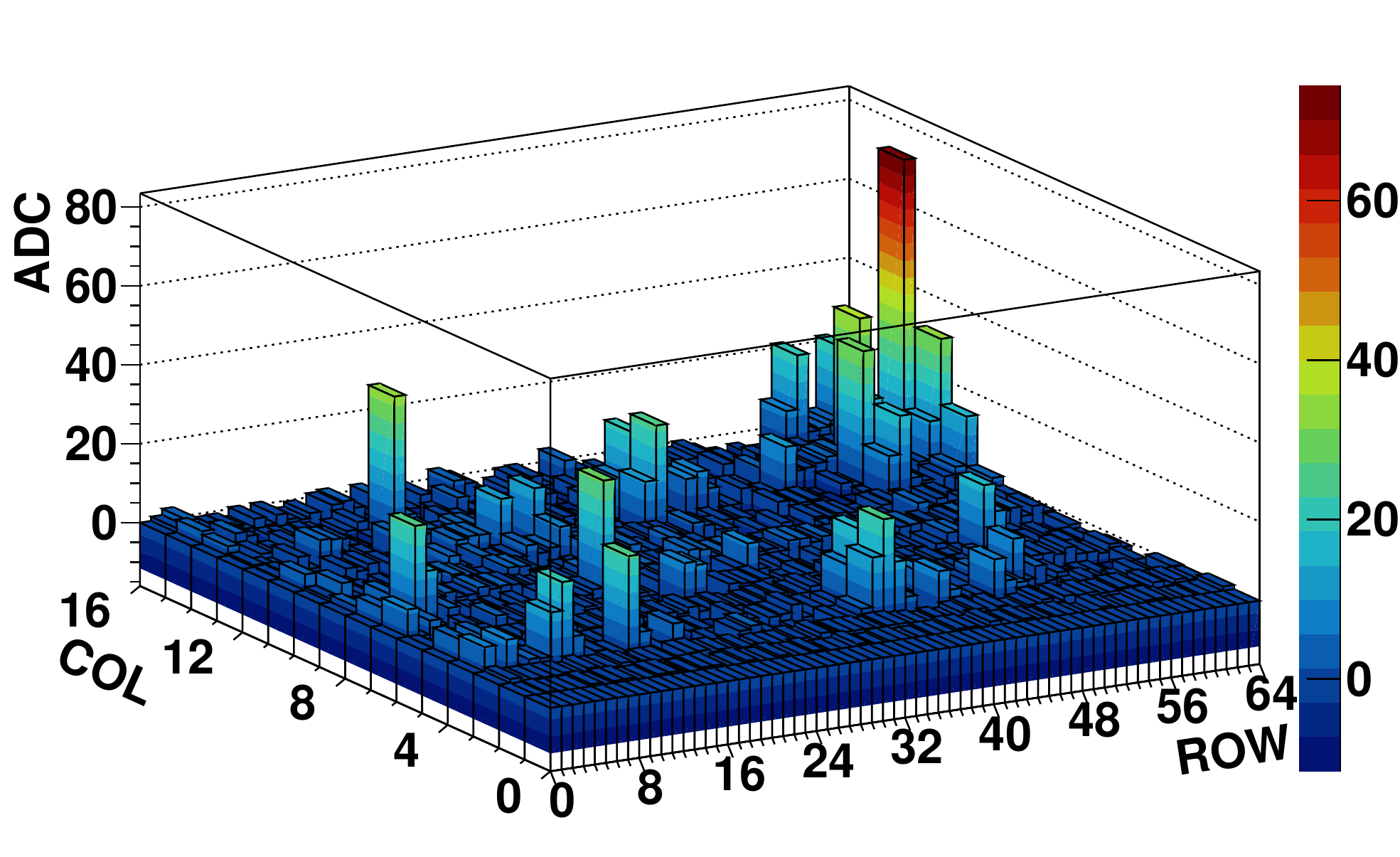}
  }%
  \subfigure[]{
    \includegraphics[width=.32\textwidth,origin=c]{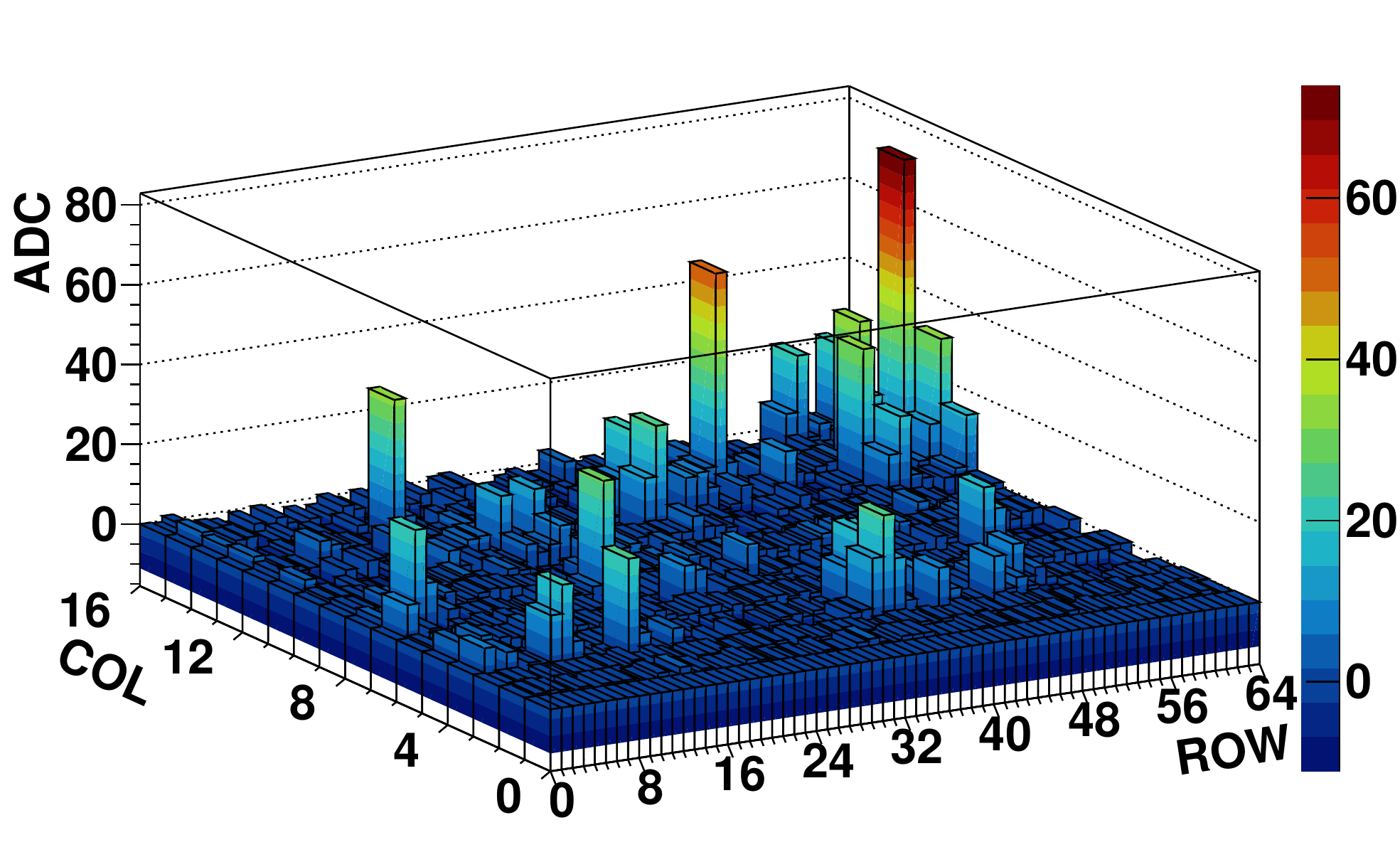}
  }%
  \subfigure[]{
    \includegraphics[width=.32\textwidth,origin=c]{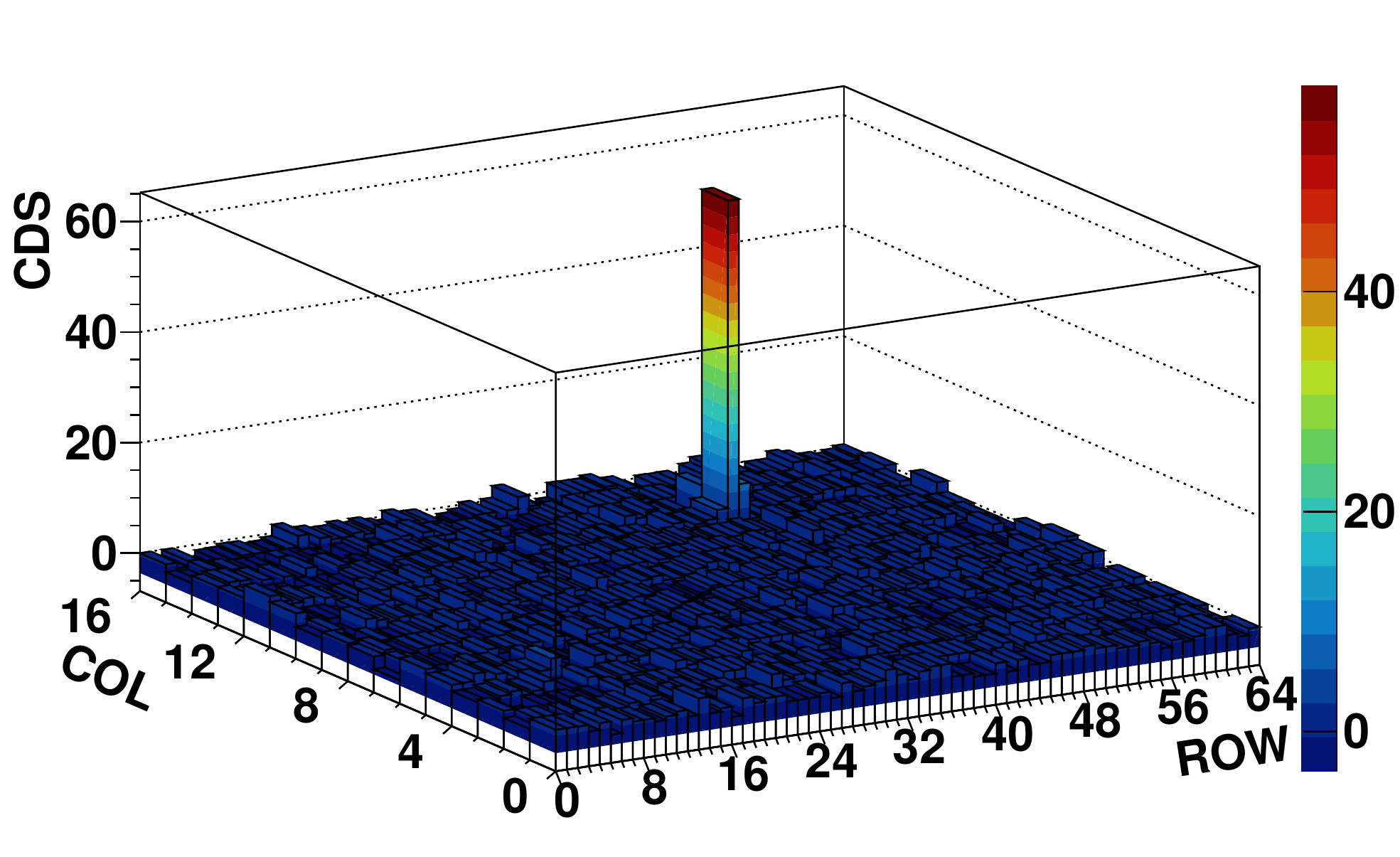}
  }

  \caption{Demonstration of the correlated double sampling (CDS)
    method: (a) the ADC output of a frame, (b) the ADC output of a
    preceding frame and (c) the result of (a) minus (b), namely CDS of
    the frame.}
  \label{fig:cds}
\end{figure}

As the readout of \supixi is in rolling shutter, the SupixDAQ takes data in
frames continuously and implements the correlated double sampling (CDS)
method. As demonstrated in \fig{cds}, the ADC distribution of a frame
is subtracted by that of a preceding frame, resulting in the CDS
distribution.
The CDS can cancel the pixel-wise fluctuation of baseline and suppress
the fixed pattern noise significantly.
Hence, the CDS value of ADC, also called ADC, will be used in the work
unless explicitly distinguished.

\begin{figure}[htb!] 
  \centering
  \includegraphics[width=\textwidth]{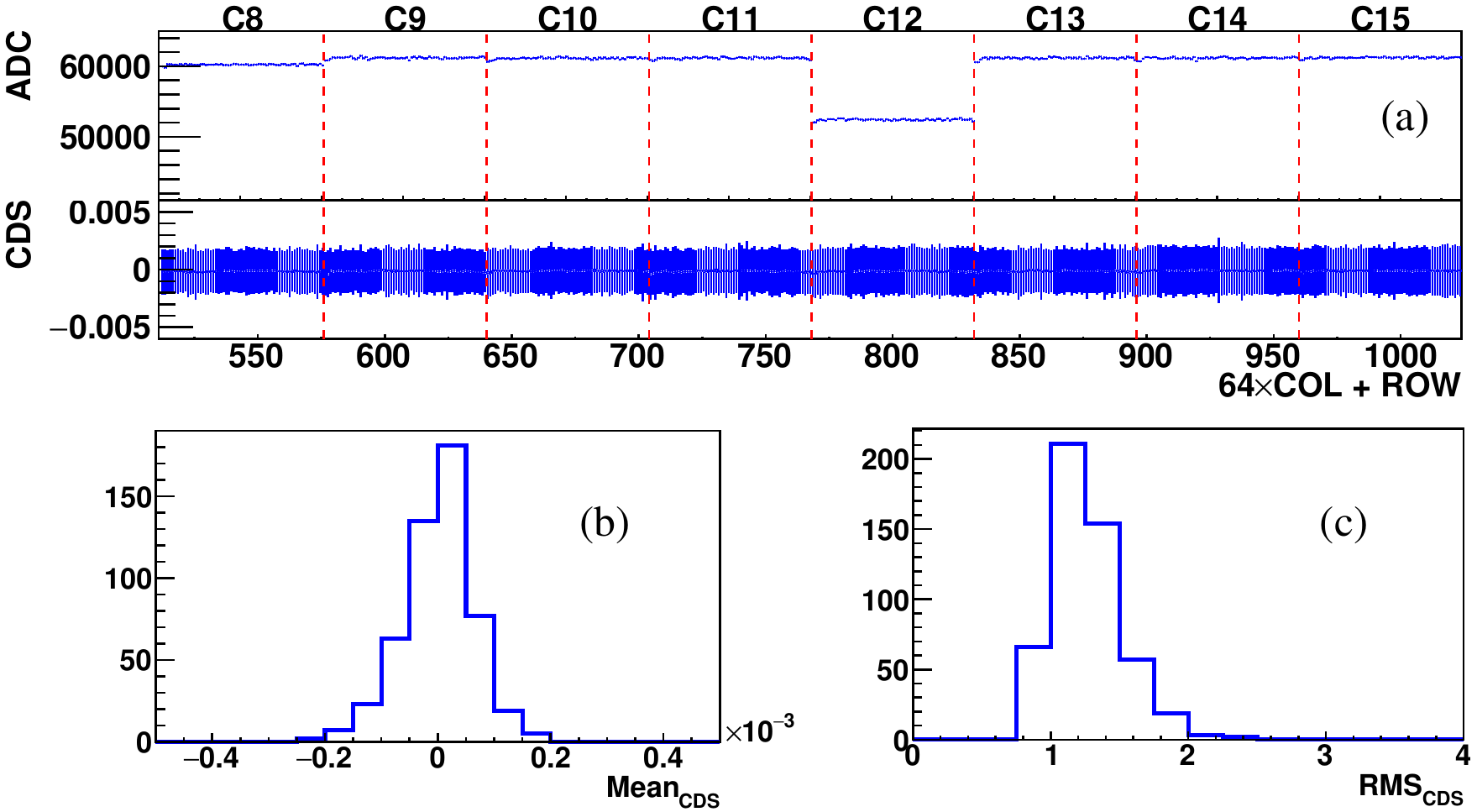}
  \caption{Results of a noise run for the sector of A0:
    (a) profiles of the mean and standard deviation of the
    pixel-by-pixel distribution of the raw ADC (up) and the ADC after
    CDS (down),
    and distributions of the CDS mean (b) and standard deviation
    (c) of pixels. Only pixels of the columns from C8 to C15 are shown.}
  \label{fig:noise}
\end{figure}

In order to calibrate noise pixel-by-pixel, a \emph{noise run} is
taken before taking data with a radioactive source, namely a
\emph{calibration run}.  As an example, the noise level before and after
CDS of each pixel in the sector of A0 is shown in \fig{noise}(a).
Both distributions of the mean and the standard deviation of the
pixel-by-pixel noise are shown in \fig{noise}(b) and (c), separately.
Values of the latter are noise calibration constants.
During the test, because the pixels of a few columns (C0, C1, C4 and C7)
did not work properly and the reconstruction of the pixel cluster
would be studied, only the pixels of columns from C8 to C15 were used.

\begin{figure}[htb!] 
  \centering
  \includegraphics[width=\textwidth]{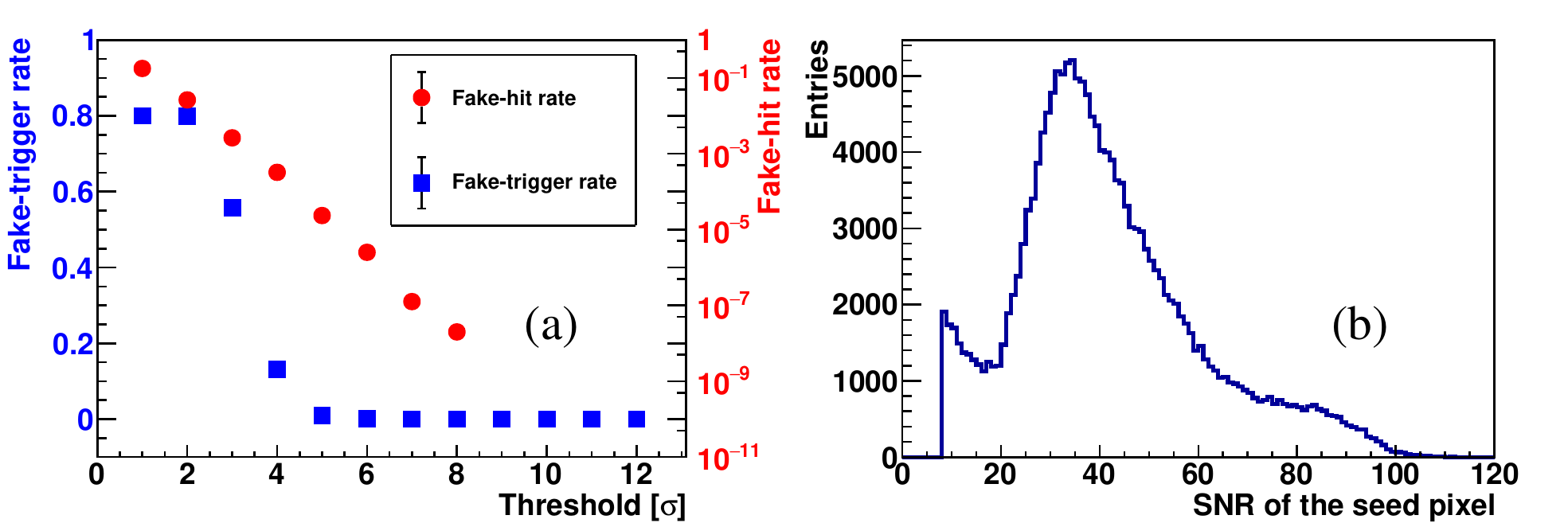}
  \caption{(a)
    The fake-trigger rate (left axis) and the fake-hit rate (right axis)
    versus the trigger threshold, and
    (b) the SNR distribution of the seed pixel.
    See the text for explanations.
  }
  \label{fig:threshold}
\end{figure}

The noise calibration constants are used in the calibration run to
define the hit pixel, with the condition of $-\text{CDS} > 8\sigma$.
A frame having at least one hit is triggered for recording for off-line
analysis.
The \emph{trigger threshold} of $8\sigma$ is chosen to suppress the
noise hit as many as possible while losing the physical hit as few as
possible.
\fig{threshold}(a) shows the fake-trigger rate and the fake-hit rate
as functions of the trigger threshold for a noise run. For the
threshold beyond $5\sigma$, the fake-trigger rate approaches zero and
the fake-hit rate decreases exponentially as the trigger threshold
increases.
\fig{threshold}(b) shows, for a calibration run, the distribution of
the signal-to-noise ratio (SNR) of the seed pixel, which has the
largest value in a frame, and the SNR is just
$-\text{CDS}/\sigma$. Obviously the cut on $8\sigma$ loses physical
hits rarely, if any. In addition, the SNR distribution of all pixels
for the same run is shown in \fig{cluster}(a).
With the chosen threshold, the rates of fake trigger and fake hit are
\SI{7(3)e-6}{trigger^{-1}} and \SI{1.8(6)e-8}{pixel^{-1}.trigger^{-1}},
respectively.

\begin{figure}[htb!] 
  \centering
  \includegraphics[width=\textwidth]{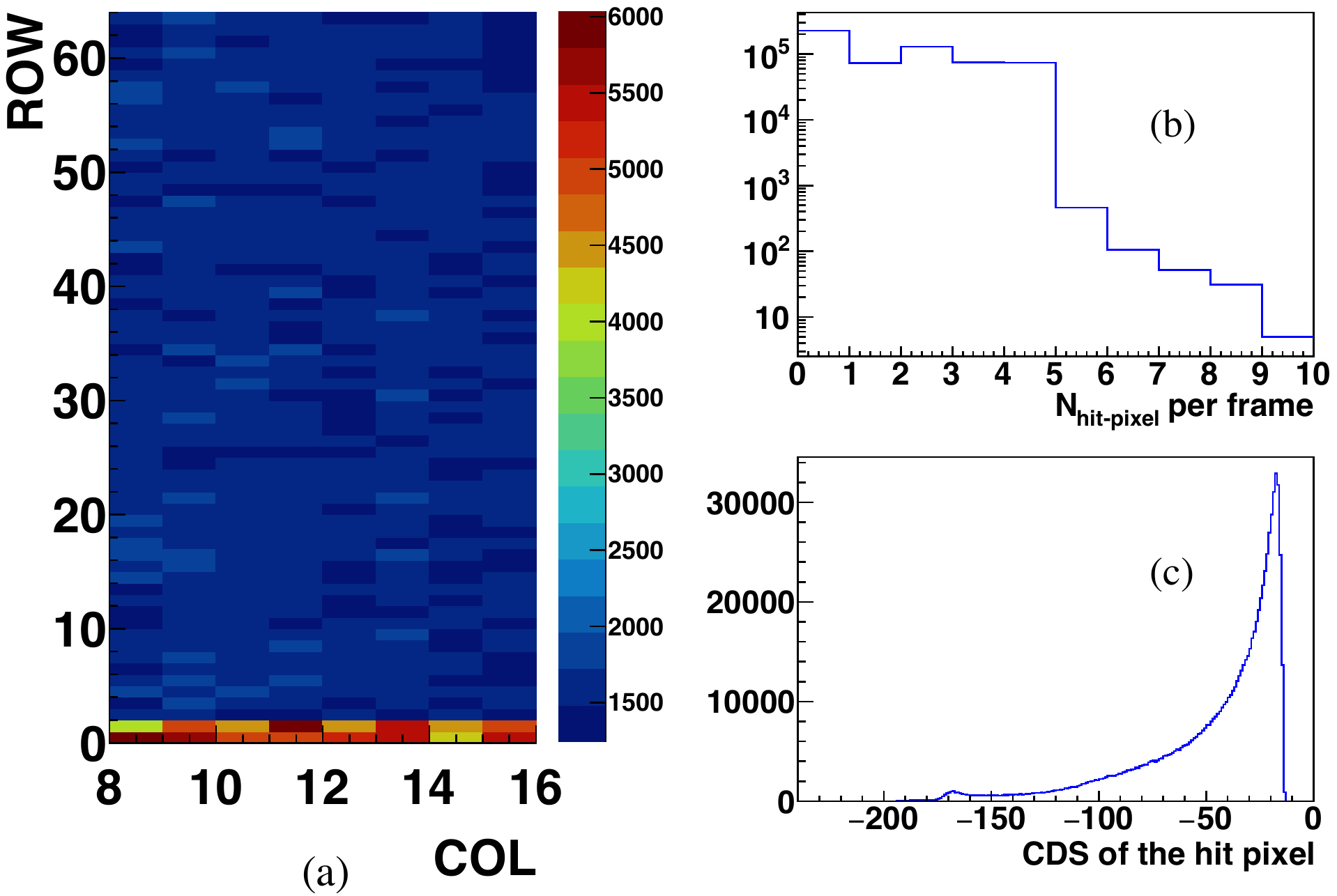}
  \caption{ Sensor performance under \X{Fe}{55} irradiation: (a) hit
    map of pixels, (b) distribution of the number of hit pixels per frame
    and (c) distribution of the ADC after CDS of the hit pixel.}
  \label{fig:hitmap}
\end{figure}

A hit map of the pixel for a sector is shown in \fig{hitmap}(a).
There are much more hits in the first two rows than in the others,
which happens in all sectors.
The cause is unknown yet, and the two rows of pixels were excluded
from searching the seed pixel.
Corresponding distributions of the number of hit pixels per frame
and the CDS of hit pixels are shown in \fig{hitmap}(b) and \fig{hitmap}(c),
respectively.
As the CDS value of the hit pixel is negative, we use its opposite
value in the following text for convenience.

\section{The chip test with  \X{Fe}{55}}\label{sect:test}

\subsection{The sensor gain calibration}
\label{sect:calib}
\begin{figure}[htb!]
  \centering
    \includegraphics[width=1.\textwidth]{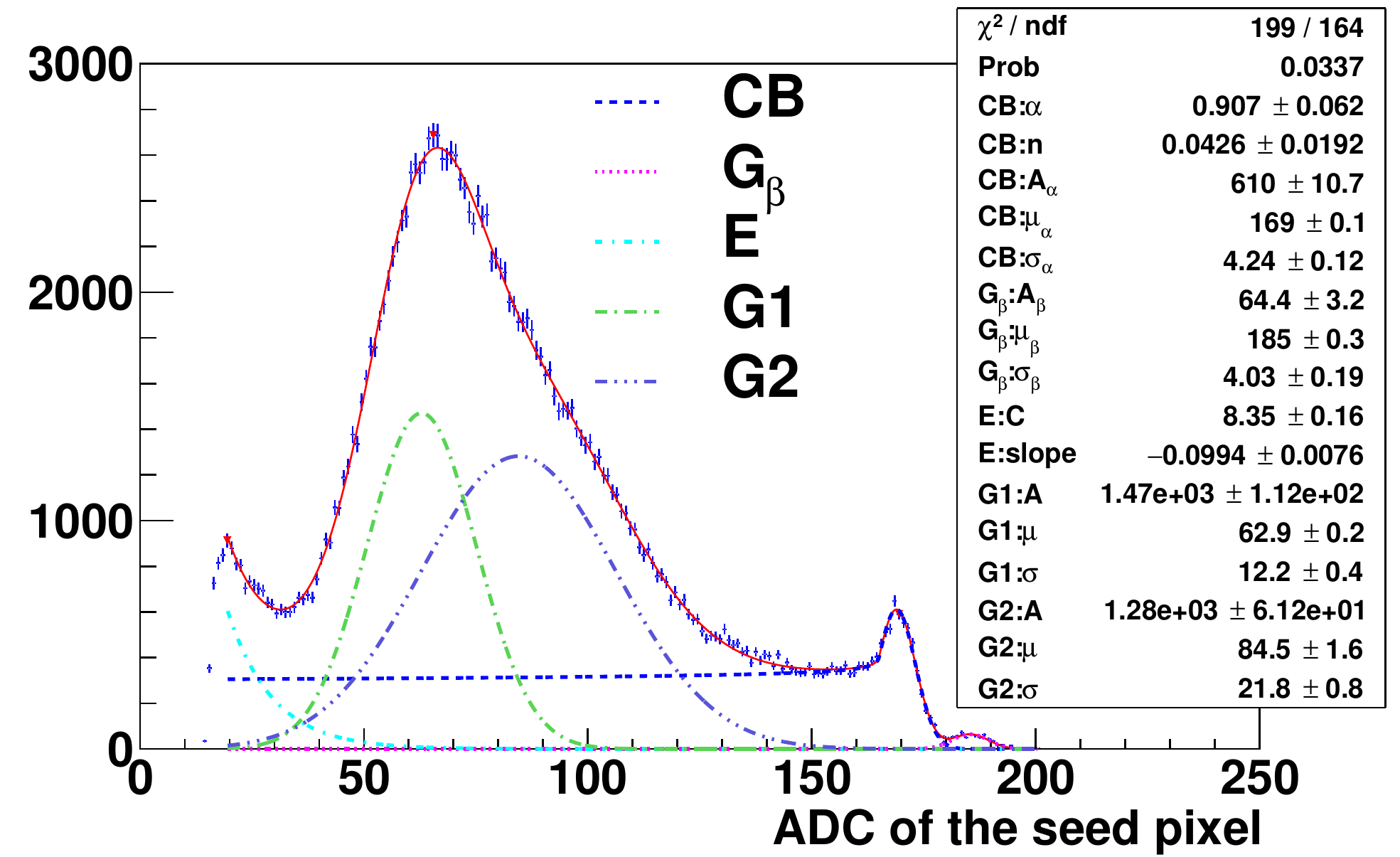}
    \caption{The distribution of the seed-pixel ADC, irradiated with
      \X{Fe}{55}, together with a full spectrum fit. See the text for
      explanations.}
    \label{fig:seed:calib}
\end{figure}

The sensor gain calibration is carried out using the soft X-rays
of \X{Fe}{55}, of which k-$\alpha$ (5.9\keV) accounts for 24.4\% and
k-$\beta$ (6.5\keV) for 2.9\% in yield. A 5.9\keV photon can generate
about 1640 electron-hole pairs when absorbed in silicon bulk, assuming
a constant value, 3.6\eV at room temperature, is required to create an
electron-hole pair~\cite{PDG}.
In most cases, the ionized electrons are collected by several pixels,
forming a pixel cluster, through thermal diffusion in the epitaxial
layer.
Specifically, the electrons will be collected by a single pixel if the photons are 
absorbed in the depleted region of that pixel,   
forming the full energy peak~\cite{FullEnergyPeak_Fe55}.
\fig{seed:calib} shows, for instance, a distribution of the seed-pixel
ADC, on which two full-energy peaks (\emph{calibration peaks}) around
169 and 185 can be observed, corresponding to the X-rays of k-$\alpha$
and k-$\beta$, respectively.

As shown in \fig{seed:calib}, the full spectrum can be fitted well with a
combination of component functions.
The addition of a crystal ball function (CB, blue dashed line) for
k-$\alpha$ peak and a Gaussian function ($\text{G}_{\beta}$) for 
k-$\beta$ peak is used to describe the calibration peaks. 
The power-law tail of the crystal ball function is justified 
considering the partial energy deposition in the depleted region.
The peak with the highest amplitude, named the \emph{collection peak},
is described by two Gaussian functions~(G1 and G2) superposed on the
crystal-ball tail.
There are two ways that ionized electrons can contribute to the
collection peak: (a) directly by an incident X-ray photon in the
vicinity of the seed pixel, and (b) by charge sharing from neighbor
pixels. Hence, the use of two Gaussians could be justified.
The Landau distribution fails to describe the collection peak because
most incident X-ray photons are fully absorbed in the epitaxial layer.
Towards the low end of the spectrum, an exponential function (E) is
added to describe the pedestal associated with the electronic noise.

The pixel gain is defined as
\begin{equation}
  \label{eq:gain}
  G = \frac{\text{ADC}_\text{peak}-\text{ADC}_\text{pedestal}}{\text{Q}},
\end{equation}
where $Q$ is for charge. Assuming the linear response, the gain can be
obtained by comparing the locations of the two calibration peaks, with no need
for pedestal subtraction.
After the gain calibration, the charge-to-voltage factor (CVF) can be
determined by
\begin{equation}
  \label{eq:cvf}
  \text{CVF} = \frac{G}{A},
\end{equation}
where $A$ is the electronics calibration factor in \eq{electronics}.
Also, the fixed pattern noise shown in \fig{noise}(c)
can be transformed into the equivalent noise charge (ENC),
\begin{equation}
  \label{eq:enc}
  \text{ENC} = \frac{\mean{\text{RMS}_\text{noise}}}{G},
\end{equation}
where $\mean{\text{RMS}_\text{noise}}$ is the mean of $\text{RMS}_\text{CDS}$
in \fig{noise}(c).
The ENC and the CVF are inversely correlated.

We use the fitted location of the collection peak, namely the most
probable value (MPV) of ADC, to characterize the charge collection
efficiency (CCE) and the SNR of the pixel for a sector,
\begin{align}
  \text{CCE}_\text{pixel}
  & = \frac{\text{ADC}_\text{MPV, seed}}{\text{ADC}_{\text{k-}\alpha, \text{seed}}},
  \\
  \text{SNR}_\text{pixel}
  & = \frac{\text{ADC}_\text{MPV, seed}}{\mean{\text{RMS}_\text{noise}}},
\end{align}
where $\text{ADC}_{\text{k-}\alpha, \text{seed}}$ is the location of
the k-$\alpha$ peak.
Results for all sectors are summarized in \tab{result}.

\subsection{The anomalous pixel output and a correction method}
\label{sect:correction}

\begin{figure}[htb!]
  \centering
  \includegraphics[width=\textwidth]{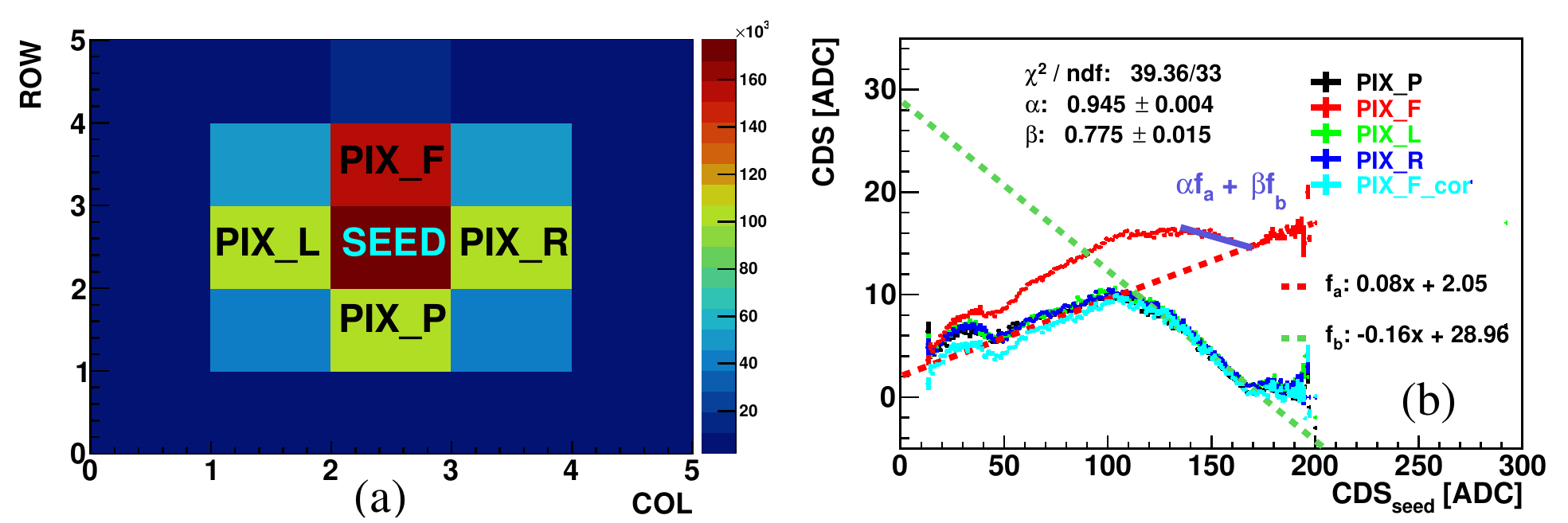}
  \caption{ (a) Distribution of the relative position of the top five
    pixels ordered in ADC within a $5\times5$ array centering at the
    seed pixel per frame,
    and (b) the ADC of each adjacent pixel and the corrected ADC of
    the anomalous pixel as a function of the seed-pixel ADC.  The tag
    PIX\_F stands for the anomalous pixel. See the text for
    explanations.  }
  \label{fig:anomaly}
\end{figure}

An anomaly was observed when we studied reconstruction for the pixel
cluster. For each frame in a $5\times5$ array centering at the seed
pixel, we figured out the top five pixels, including the seed pixel,
based on the ordering of ADC.
\fig{anomaly}(a) shows the relative position distribution of the
top five pixels. Among four adjacent pixels, the frequency of the pixel in
the same column as the seed but being read out immediately after the
seed, tagged as PIX\_F, is anomalously high. 
In \fig{anomaly}(b), the ADC of each adjacent pixel is drawn as a
function of the seed-pixel ADC, and the anomaly is
significant. Particularly, in the ADC region greater than 170, where 
the seed pixel absorbs mostly all ionized charges, the ADC of the 
three normal pixels fluctuates around zero. In comparison, the 
ADC of the anomalous pixel increases linearly as the seed ADC 
increases. The anomaly might be related to the rolling shutter 
readout, but remains unknown yet.

An empirical method was developed to correct the output of the
anomalous pixel, that is also illustrated in \fig{anomaly}(b). A
linear function, $f_a(x)$, where $x$ stands for ADC of the seed pixel,
is obtained via fitting the anomalous curve in the range of
$x>170$. Another linear function, $f_b(x)$, is obtained via fitting
the normal curves in the range of $130 < x < 170$. Then a linear
combination of the two functions, $\alpha f_a + \beta f_b$, is fitted
against the anomalous curve in the same fit range as that of $f_b$ to
determine the coefficients $\alpha$ and $\beta$. Finally, the anomaly
is corrected with the equation
\begin{equation}
  \text{PIX\_F\_cor} = \frac{\text{PIX\_F}(x)-\alpha f_a(x)}{\beta},
\end{equation}
where PIX\_F and PIX\_F\_cor are the original and corrected ADCs of the
anomalous pixel, respectively. As shown in \fig{anomaly}(b), the
correction (cyan curve) is reasonable,
but it is somehow underestimated in the range of
$\text{CDS}_\text{seed} < 100\,\si{ADC}$. The effect of the bias to
the pixel clustering is negligible based on further study.

\subsection{Reconstruction of the pixel cluster}
\label{sect:cluster}
Most X-ray photons of \X{Fe}{55} deposit their energy in the
epitaxial layer and the ionized electrons are collected by a cluster
of pixels. Hence, the clustering algorithm as well as the property of
the cluster was studied. We tried several methods to reconstruct the
cluster of pixels. Their performance was similar and one of them is
reported here.

\begin{figure}[htb!]
  \centering
  \includegraphics[width=\textwidth]{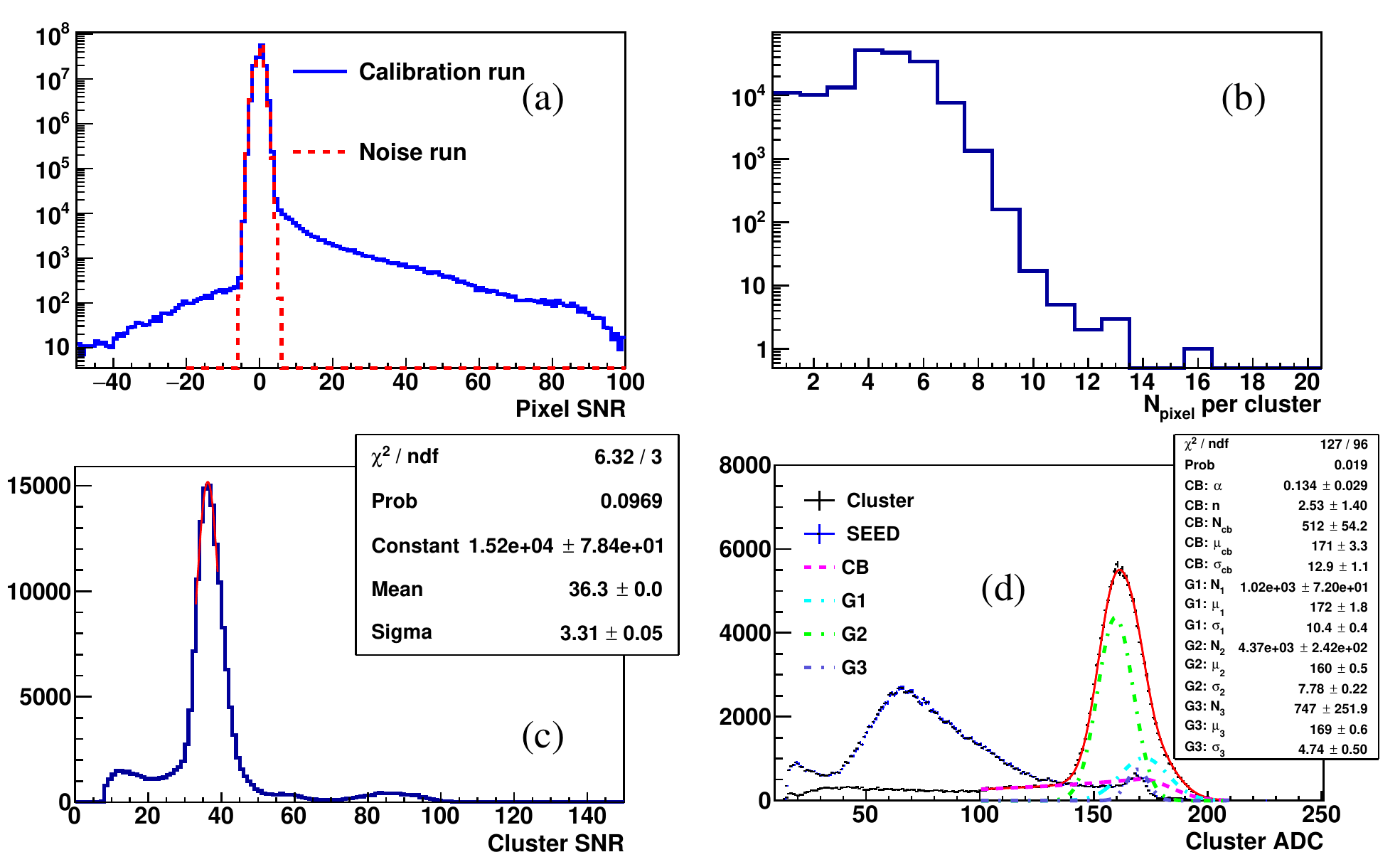}
  \caption{Reconstruction for the cluster of pixel:
    (a) distributions of the pixel SNR for a noise run and a
    calibration run,
    (b) distribution of the number of pixels per cluster,
    (c) distribution of the cluster SNR,
    and (d) distribution of the cluster ADC and a fit for the
    distribution, together with the distribution of the seed-pixel ADC
    for comparison.
    See the text for explanations.
  }
  \label{fig:cluster}
\end{figure}

% clustering method and the effect of threshold
A cluster is defined by a group of adjacent fired pixels, which have
signals above a \emph{clustering threshold}.
The cluster reconstruction starts with a seed pixel, selects neighboring
fired pixels including diagonal ones, and repeats the selection for
newly selected pixels until there are no fired neighbors anymore or
reaching an edge of the pixel array.
% threshold
The clustering threshold for the fired pixel is based on the SNR of
individual pixels.
% , defined previously in discussing the trigger threshold.
%
As indicated in \fig{cluster}(a), the threshold is set to $3.1$,
corresponding to less than one per mil of noise pixels leaked into
clustering pixels.
% cluster properties
The number of pixels and the SNR of the reconstructed cluster are
shown in \fig{cluster}(b) and (c), respectively. The cluster SNR is
evaluated as
\begin{equation}
  \text{SNR}_\text{cluster}
  = \frac{\sum_i \text{ADC}_i}{\sqrt{\sum_i \sigma_i^2}},
\end{equation}
where $\text{ADC}_i$ and $\sigma_i$ are the ADC and the noise of the
$i$-th pixel.  The numerator is just the ADC of a cluster.  Most of
the clusters consist of 4 or 5 pixels and have the SNR greater than 25.
On the SNR distribution, there is a small peak at about 85, which
attributes mainly to the single-pixel clusters.

The distribution of the cluster ADC is shown in \fig{cluster}(d),
together with that of the seed pixel for comparison.
The distribution can be fitted well with a few Gaussian functions
superposed with a crystal ball function.  The peak of the distribution
shifts toward the left with regard to the k-$\alpha$ peak.
The shift, indeed, depends on the clustering threshold, which is a
compromise between the noise pixel and the physically fired pixel.
We can use the location of the peak to define the charge collection
efficiency of the cluster,
\begin{equation}
  \text{CCE}_\text{cluster}
  = \frac{\text{ADC}_\text{MPV, cluster}}{\text{ADC}_{\text{k-}\alpha, \text{seed}}}.
\end{equation}
Both of the $\text{SNR}_\text{cluster}$ and
$\text{CCE}_\text{cluster}$ can characterize the performance of the sensor
under test. Results are summarized in \tab{result}, where the MPV of
$\text{SNR}_\text{cluster}$, as shown in \fig{cluster}(c), is
reported.

\section{Test results}\label{sect:results}

All available sectors have been tested with the previously described
procedure. Test results as well as geometrical parameters of the
sensor are listed in \tab{result}, where the pixel-wise quantities are
defined in \sect{calib} and the cluster-wise quantities are defined in
\sect{cluster}.

% gain calibration G = ADC/Q
% electronics calib. A = ADC/V
% CVF = G/A
% ENC = sigma_ADC / G [e]
% ----------------------------------------------------------------------

%
For all sectors, the CVFs and the ENCs of the pixels are consistent within
uncertainties.  
The means of pixel-by-pixel ENC distributions in each
sector are all below \SI{14}{e^-}, and have the RMS spreads of about
40\%.  Taking A0 as a reference, variations of the means among sectors
are within 10\%, less than the uncertainties.

As for the pixel-wise performance of charge collection, the SNRs are
consistent within uncertainties, whereas the CCEs, and the means of 
SNRs exhibit geometry dependence similarly.
The sectors of A2 and A0 have almost the same geometry configuration
except that the x-pitch of A2 is four times that of A0. The CCE of A2
decreases about 19\% of that of A0, while the mean of SNR decreases
about 6\% only.
The sectors of A2, A5 and A8, consisting of the largest pixels, have the
same pixel pitches and different areas of the diode surface and
footprint.
Among them, the CCE and SNR of A8 are the best, very close to those of
A0, whereas those of A5 are the worst.
It implies that not only the absolute size of the diode surface but also
the surface to footprint ratio influences the performance.
The sectors of A7 and A8 have similar diode dimensions, but the x-pitch
of A7 is half of A8's. In addition, the diodes of A8 have a staggered
arrangement, therefore the diode distribution density is half of that
of A7 evenly. That explains why A7, due to charge sharing, has  
smaller CCE than A8 does.
The observations confirm our previous simulation for the sensor
design, as shown in Figure 3 and 4 in \refcite{supix1tcad}.

For the cluster-wise performance, all sectors perform well, having
CCEs and SNRs greater than 90\% and 30, respectively.
Again the SNRs are consistent within uncertainties, and their means
depend weakly on geometrical configuration.
On the other hand, the CCEs are almost geometry independent, within
about 5\% variation referring to the CCE of A0, and A7 performs the
best.

\begin{table}[htb!]
  \caption{Geometrical parameters of the \supixi sensor and test
    results with a \X{Fe}{55} radioactive source.
    CVF is for the charge-to-voltage factor, ENC is for the equivalent
    noise charge, CCE is for the charge collection efficiency and SNR
    is for the signal-to-noise ratio.
    The uncertainties of the CCEs are on the most probable values,
    while the uncertainties of the others are of the distribution.
    See the text for more explanations. }
  \label{tab:result}
  \centering
  \input{table_result_MR}
\end{table}

\section{Conclusions}
\label{sect:conclusion}

%%% What we did.........................................................

Using a radioactive source of \X{Fe}{55}, we have characterized the
\supixi pixel sensor implemented in a \SI{180}{nm} CMOS imaging
process.
The sensor, consisting of sectors with different pixel configurations,
is aimed to investigate detection performance of the large pixel,
meeting requirements of a proposed pixelated silicon tracker at CEPC.
%
%All available sectors have effective CCEs and SNRs.
%
The full energy peaks corresponding to the k-$\alpha$ and k-$\beta$
X-rays of \X{Fe}{55} have been observed and are used to calibrate the
pixel gain and to obtain the pixel-wise CVF, ENC, CCE and SNR.
A preliminary algorithm to reconstruct the signal cluster of pixels
has been developed and the cluster-wise CCE and SNR are
determined. During the study of reconstruction, an anomaly in the pixel
output
%, which might relate to the rolling-shutter readout,
is observed and an empirical correction method has been developed.

%%% What results we obtained............................................

%% noise
With our chosen threshold for data taking, the fake-hit rate is
in the order of \SI{e-8}{pixel^{-1}.trigger^{-1}}.
The pixels of all available sectors have low noise, with all ENCs
being about \SI{13}{e^-} and almost the same within the uncertainty of
about \SI{5}{e^-}.
%
%% charge collection
For different sectors, the pixel-wise CCEs depend on the geometry
configuration significantly, whereas the cluster-wise CCEs do not and
are greater than 90\% for all.
%
%
%% SNR
The SNRs of both pixel and cluster in all sectors are high, mostly
above 30, and are consistent within uncertainties.

%%% What conclusions we draw............................................

%% A8 is good for CEPC
Our test results demonstrate that large pixels, like those in the sector
of A8, satisfy typical requirements of pixel detectors in the sense of
noise and charge collection
efficiency. %, and are suitable for tracking at CEPC.
%
%% CVF and ENC not constrain...
They are practical for tracking at CEPC, since CVF and ENC do not
constrain pixel performance at the large sizes.
%
%% optimization
In addition, the pixel performance of charge collection depends not
only on the pitch size but also on the diode design, including the
areas of surface and footprint as well as the arrangement of position.
It would be potentially possible to explore the design of even larger
pixels, optimized towards specific requirements of different
detectors.

%%% end of article______________________________________________________

\acknowledgments

This work has been supported by the National Natural Science 
Foundation of China(U1232202, U2032203 and 12075142), the 
Ministry of Science and Technology of China (2018YFA0404302) 
and Shandong Provincial Natural Science Foundation (ZR2020MA102).

%%% References
\bibliographystyle{JHEP_wm}
\bibliography{supix1}
 
\end{document}

%% file: table_result_MR.tex
  \begin{tabular}{c|c|c|c|c|c|c}
    \hline
    \multicolumn{2}{c|}{Sector}& A0 & A2 & A5 & A7 & A8\\
        \hline\hline
        \multicolumn{2}{c|}{Sensitive Area x (mm)} & 0.3 & 1.3 & 1.3 & 0.7 & 1.3\\
        \hline
         \multicolumn{2}{c|}{Sensitive Area y (mm)} & \multicolumn{5}{c}{1.3} \\
        \hline
        \multicolumn{2}{c|}{Pixel Pitch x ($\mu$m)} & 21 & 84 & 84 & 42 & 84 \\
        \hline
        \multicolumn{2}{c|}{Pixel Pitch y ($\mu$m)} & \multicolumn{5}{c}{21} \\
        \hline
        \multicolumn{2}{c|}{Diode Surface ($\mu$m$^2$)} & 8 & 8 & 12 & 20 & 20  \\
        \hline
        \multicolumn{2}{c|}{Diode Footprint ($\mu$m$^2$)} & 11 & 11 & 18 & 44 & 50\\
        \hline\hline
        \multirow{4}{*}{Pixel} & CVF (\si{\mu V/e})& 13$\pm$5 & 14$\pm$5 & 12$\pm$4 & 14$\pm$4 & 13$\pm$5\\
        						\cline{2-7}
        						  & ENC (e) & 13$\pm$5 & 12$\pm$5 & 13$\pm$5 & 12$\pm$4 & 12$\pm$5\\
						  \cline{2-7}
						  & CCE (\%) &  39.1$\pm$1.0 & 31.7$\pm$0.7  & 27.5$\pm$0.6  & 30.4$\pm$0.8  & 37.2$\pm$0.7 \\
						  \cline{2-7}
						%%& FWHM (ADC) & 52.7  & 44.9  & 51.8  & 64.6  & 45.3 \\
						 %% & $\sigma_\text{Q}$ (e)& 532.4$\pm$188.2 & 408.0$\pm$144.6 & 539.3$\pm$191.0 & 603.5$\pm$163.6 & 439.6$\pm$162.2\\ 
						%  & $\sigma_\text{Q}$ (\%) & 79.3$\pm$2.0 & 69.7$\pm$1.6 & 113.7$\pm$2.6 & 115.9$\pm$2.8 & 66.3$\pm$1.2\\
						 % \cline{2-7}
						  & SNR & 52$\pm$11  & 49$\pm$10  & 36$\pm$8  & 42$\pm$9  & 53$\pm$11 \\
						  
	\hline
%	\multirow{2}{*}{Cluster} & CCE (\%) & 95.6$\pm$6.2 & 90.3$\pm$6.5 &  93.8$\pm$6.5 & 96.8$\pm$5.4 & 92.4$\pm$6.1 \\
%							\cline{2-7}
	\multirow{2}{*}{Cluster} & CCE (\%) & 95.6$\pm$0.5 & 90.3$\pm$0.4 &  93.8$\pm$0.5 & 96.8$\pm$0.4 & 92.4$\pm$0.6 \\
							\cline{2-7}
							%&$\sigma_\text{Q}$ (e) & 236.8$\pm$83.7 & 277.5$\pm$98.4 & 270.5$\pm$95.8 & 261.5$\pm$11.6 & 264.2$\pm$97.5\\
						       	%& $\sigma_\text{Q}$(\%) & 14.5$\pm$0.7& 16.6$\pm$0.8 & 16.8$\pm$0.8 & 13.3$\pm$0.7 & 16.0$\pm$0.8\\
							%\cline{2-7}
						        & SNR & 36$\pm$3 & 33$\pm$2 & 32$\pm$2 & 36$\pm$3 & 36$\pm$3  \\
						       
	\hline
  \end{tabular}